\documentclass[final,5p,times,twocolumn]{elsarticle}

\usepackage{amsmath}
\usepackage{bm}
\usepackage{graphicx}% Include figure files
\usepackage{xcolor}
\usepackage{multirow}
\usepackage{booktabs}
\usepackage{orcidlink}
\usepackage{xspace}
\usepackage{hyperref}% add hypertext capabilities

\newcommand{\Brms}{B_{\mathrm{rms}}}
\newcommand{\Bsat}{B_{\mathrm{sat}}}
\newcommand{\vrms}{v_{\mathrm{rms}}}

\newcommand{\teddy}{t_{\mathrm{eddy}}}
\newcommand{\tcool}{t_{\mathrm{cool}}}
\newcommand{\tdyn}{t_{\mathrm{dyn}}}

\newcommand{\Rm}{R_{\mathrm{m}}}
\newcommand{\Rmcrit}{R_{\mathrm{m,crit}}}
\newcommand{\Pm}{P_{\mathrm{m}}}

\newcommand{\Emag}{E_{\mathrm{mag}}}
\newcommand{\Ekin}{E_{\mathrm{kin}}}

\newcommand{\fsat}{f_{\mathrm{sat}}}

\newcommand{\Gammac}{\Gamma}
\newcommand{\Gammam}{\Gamma_{\mathrm{m}}}
\newcommand{\Tmelt}{T_{\mathrm{melt}}}
\newcommand{\Hp}{H_{\mathrm{P}}}

\newcommand{\flash}{\textsc{flash}\xspace}

\journal{Journal of High Energy Astrophysics}

\begin{document}

\begin{frontmatter}

\title{A post-hypercritical accretion small-scale dynamo in newborn neutron stars}

\author[oan]{David F.~Bambague \orcidlink{0009-0004-7604-8796}}
\ead{dbambague@unal.edu.co}

\author[fisunal]{Cristian G.~Bernal \orcidlink{0000-0003-0370-7397}}
\ead{cribernal@unal.edu.co}

\author[icranet,icra,icranetferrara,unife,inafaps]{J. A. Rueda \orcidlink{0000-0003-4904-0014}}
\ead{jorge.rueda@icra.it}

\address[oan]{Observatorio Astron\'omico Nacional, Universidad Nacional de Colombia, Bogot\'a, Colombia}

\address[fisunal]{Departamento de F\'isica, Facultad de Ciencias, Universidad Nacional de Colombia, Bogot\'a, Colombia}

\address[unife]{Dipartimento di Fisica e Scienze della Terra, Universit\`a degli Studi di Ferrara, Via Saragat 1, 44122 Ferrara, Italy}
\address[icranet]{ICRANet, Piazza della Repubblica 10, 65122 Pescara, Italy}

\address[icra]{ICRA, Dipartimento di Fisica, Sapienza Università di Roma, Piazzale Aldo Moro 5, I-00185 Roma, Italy}

\address[icranetferrara]{ICRANet-Ferrara, Dipartimento di Fisica e Scienze della Terra, Universit\`a degli Studi di Ferrara, Via Saragat 1, 44122 Ferrara, Italy}

\address[inafaps]{INAF, Istituto di Astrofisica e Planetologia Spaziali, Via Fosso del Cavaliere 100, I-00133 Rome, Italy}

\begin{abstract}
Hypercritical fallback accretion can advect the surface magnetic field of a newborn neutron star into the newly accreted outer layers. Before this material joins the solid crust and enters the Hall-Ohmic regime, part of it may remain hot, dense, and liquid. This short-lived fluid stage may support turbulent magnetic amplification. We test whether a small-scale dynamo (SSD) can be activated under thermodynamic conditions of the liquid post-hypercritical layer. We quantify the associated growth rates, amplification factors, and saturation levels, adopting a matter density $\rho_0 = 10^{10}$ g cm$^{-3}$ and a temperature $T_0 = 2\times10^9$ K. We perform six local 3D resistive MHD simulations with \flash~4.7 in a $(100\,\text{m})^3$ periodic domain with externally forced subsonic turbulence. The reference model uses the Helmholtz equation of state (EOS) and neutrino cooling. Three runs with a $128^3$ mesh resolution vary the magnetic Reynolds number ($\Rm \sim 700$--$3700$), two control runs isolate the effect of the EOS and neutrino cooling, and one $256^3$ resolution run tests the robustness of the reference case. For $\Rm \sim 700$--$3700$, the magnetic field grows exponentially from $B_0 = 10^{12}$ G and saturates at $\Bsat \sim 3$--$7\times10^{13}$ G within millisecond timescales. The saturated magnetic energy remains sub-equipartition, with a magnetic-to-kinetic energy ratio $\fsat = \Emag/\Ekin \approx 0.2$--$0.3$, consistent with an SSD behavior at magnetic Prandtl number $\Pm \sim 1$. The results of the $128^3$ and $256^3$ reference runs agree to within a few percent. Neutrino cooling does not affect the dynamics over the simulated time, and the choice of EOS changes the dynamo metrics only weakly in the subsonic regime explored here. These simulations show that a forced local SSD can operate efficiently in a liquid post-hypercritical accretion layer. Further simulations in a stratified, decaying post-fallback flow over the lifetime and energy reservoir of the inherited turbulence are needed for assessing the global model of magnetic reemergence and amplification.

\end{abstract}


\end{frontmatter}
\noindent Accepted for publication in the Journal of High Energy Astrophysics (JHEAP).

% =======================================================
% ====INTRODUCTION=======================================
% =======================================================

\section{Introduction}
\label{sec:intro}

Hypercritical fallback accretion can substantially modify the magnetic-field structure of a newborn neutron star (NS). If the fallback rate is sufficiently high, the ram pressure of the accreting material overwhelms the magnetic pressure and advects the surface magnetic field into the newly deposited outer layers \cite{Chevalier1989,HouckChevalier1991,MuslimovPage1995,1996ApJ...460..801F}. This process, commonly referred to as magnetic-field burial or submergence, has been investigated through analytic models and numerical simulations \cite{GeppertPageZannias1999,Bernal2013,TorresForne2016} and is frequently invoked to explain young NSs whose inferred external dipole fields appear much weaker than the internal magnetic fields suggested by their thermal or high-energy emission properties \cite{BernalFraija2016,Fraija2014,Fraija2015,Fraija2018,Fraija2019}.

Most studies of buried magnetic fields focus on either of two evolutionary stages. The first is the field-generation phase inside the proto-NS, where convection, differential rotation, and magnetorotational instabilities can amplify magnetic fields shortly after core collapse \cite{DuncanThompson1992,ThompsonDuncan1993,Akiyama2003,Obergaulinger2009,Mosta2015,Raynaud2020,ReboulSalze2021}. The second is the long-term evolution of the solid crust, where Hall drift and Ohmic dissipation govern the transport and reemergence of magnetic flux \cite{GoldreichReisenegger1992,CummingArrasZweibel2004,PonsGeppert2007,PonsMirallesGeppert2009,Vigano2013,GourgouliatosCumming2014,Gourgouliatos2016,PonsVigano2019,IgoshevPopov2021}. Little attention has been devoted to the intermediate stage between these two regimes.

After hypercritical accretion ceases, the freshly deposited material is hot, dense, and liquid (see Sec. \ref{sec:crystallization}) before eventually crystallizing and entering the Hall-Ohmic regime. During this transient phase, the buried magnetic field is embedded in a conducting fluid that may still contain turbulent motions inherited from the fallback flow. This naturally raises the question of whether magnetic amplification can continue after field burial and before crust crystallization.

A possible mechanism is the small-scale turbulent dynamo (SSD), in which random three-dimensional motions stretch and fold magnetic field lines, leading to exponential growth of magnetic energy once the magnetic Reynolds number exceeds a critical threshold \cite{Kazantsev1968,BrandenburgSubramanian2005,Schekochihin2004ApJ,Schekochihin2004PRL}. SSD action has been extensively studied in laboratory plasmas, the interstellar medium, and numerical simulations of compressible turbulence \cite{HaugenBrandenburgDobler2004,Federrath2011,Schober2012,SetaFederrath2020,SetaFederrath2021,SetaFederrath2022,Kriel2022}. However, its possible operation in the liquid post-hypercritical accretion layer of a newborn NS has received little attention. To our knowledge, the occurrence of SSD amplification during this pre-crystallization stage has not been investigated using dedicated MHD simulations.

This possibility is particularly interesting in the context of Central Compact Objects (CCOs) and low-field magnetars \cite{DeLuca2008,GotthelfHalpern2005}. Timing observations of several CCOs indicate external dipole fields of only $10^{10}$--$10^{11}$ G \cite{HalpernGotthelf2010,Gotthelf2013}, while pulse-profile modeling and thermal anisotropies often require substantially stronger internal magnetic structures \cite{ShabaltasLai2012,Bogdanov2014,Luo2015,Igoshev2021CCO}. Likewise, low-field magnetars display magnetar-like activity despite relatively modest inferred dipole fields \cite{Rea2010,Rea2012,Rea2014}. These systems motivate scenarios in which a significant fraction of the magnetic energy remains hidden beneath the stellar surface during the early stages of evolution.

The purpose of this work is to investigate whether a local SSD can operate under the thermodynamic conditions of the liquid post-hypercritical accretion layer. We perform three-dimensional resistive MHD simulations using FLASH 4.7 \cite{Fryxell2000} in a locally periodic domain with externally driven subsonic turbulence. Our goal is not to model the global fallback flow or subsequent crustal evolution, but to determine the local turbulent magnetic amplification under the physical conditions expected in the post-fallback liquid layer and to quantify the associated growth rates and saturation levels.

The paper is organized as follows. In Sec.~\ref{sec:crystallization}, we estimate the viability of a liquid post-hypercritical window and compare cooling, crystallization, and dynamo timescales. Section~\ref{sec:setup} describes the numerical model, including the MHD equations, microphysics, forcing prescription, resistivity, and run matrix. Section~\ref{sec:results} presents the magnetic-energy growth, saturation behavior, dependence on magnetic Reynolds number, control runs, and resolution check. In  Sec.~\ref{sec:limitations}, we discuss the limitations of the present model and numerical setup. Section~\ref{sec:conclusions} summarizes the main conclusions. 

% ===========================================================
\section{Is the post-hypercritical accretion layer liquid?}
\label{sec:crystallization}
% ============================================================

We start by evaluating whether the material in the local computational domain can be treated as a fluid. Three cooling regimes must be distinguished in the early evolution of a newborn NS. During the proto-NS phase ($\sim10^{-3}$--$10^{1}\,\mathrm{s}$), neutrinos are trapped in the hot interior and the cooling is controlled by diffusion from the core to the neutrinosphere \cite{BurrowsLattimer1986,Pons1999}. After the crust has crystallized, the magnetic and thermal evolution is governed by Hall drift, Ohmic dissipation, and heat transport in a solid medium \cite{Vigano2013,PonsVigano2019}. The conditions considered here belong to the intermediate, neutrino-transparent envelope regime, in which the outer layers at $\rho\sim10^{8}$--$10^{11}\,\mathrm{g\,cm^{-3}}$ cool through local volumetric neutrino emissivity, mainly pair annihilation at high temperature and plasmon decay at lower temperature \cite{Negreiros2012,Negreiros2020,Potekhin2015}. We show that our fiducial state, $\rho_0 = 10^{10}\,\mathrm{g\,cm^{-3}}$ and $T_0 = 2\times10^9\,\mathrm{K}$, lies in this neutrino-transparent, non-crystallized regime.

\begin{figure}[!ht]
\centering
\includegraphics[width=\hsize]{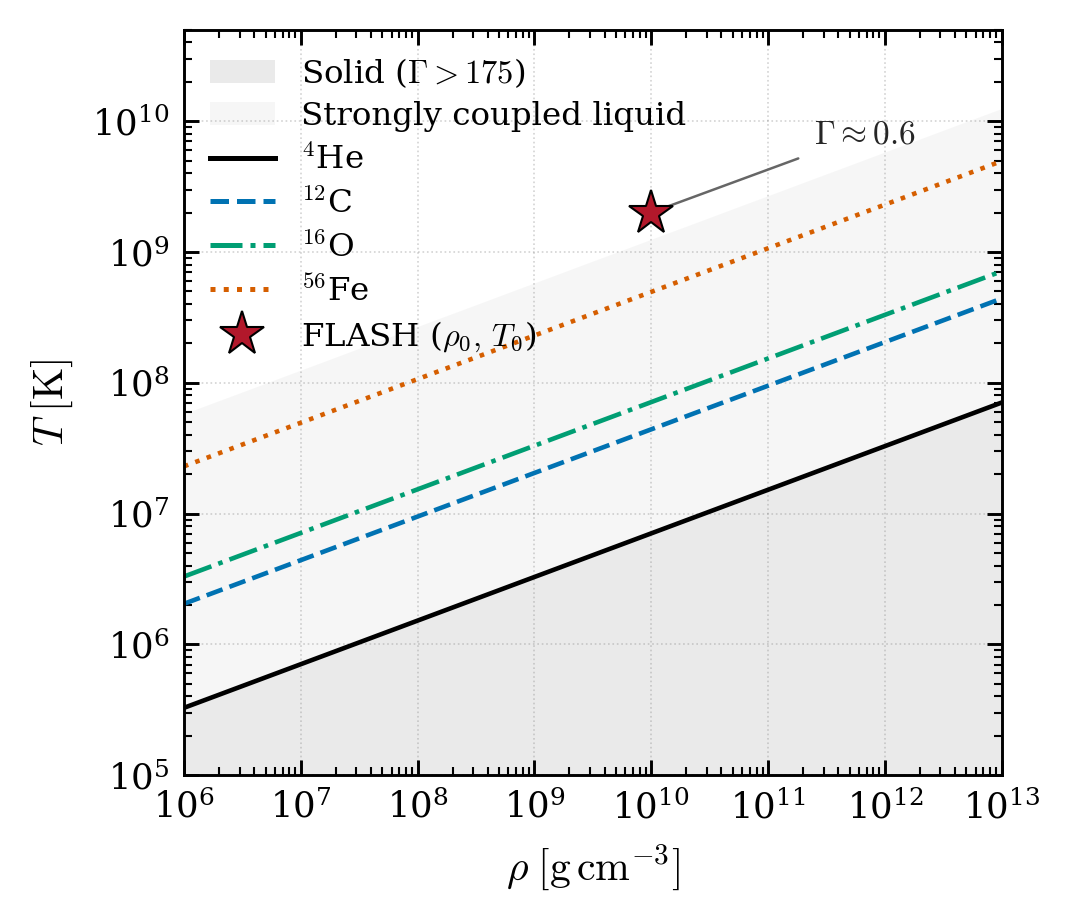}
\caption{$\rho$--$T$ phase diagram for the local post-hypercritical accretion conditions considered in this work. Curves show the melting temperature
$\Tmelt(\rho)$ from Eq.~(\ref{eq:Tmelt}) for representative compositions.
The shaded regions mark the solid phase ($\Gammac>175$) and the strongly coupled liquid regime ($1<\Gammac<175$) for He$^4$. The star marks the fiducial simulation point, $\rho_0=10^{10}\,\mathrm{g\,cm^{-3}}$ and
$T_0=2\times10^9\,\mathrm{K}$, for which
$\Gammac\simeq0.62$. The point lies well above the melting curves; even for Fe$^{56}$, $T_0$ is several times larger than $\Tmelt$.}
\label{fig:phase_diagram}
\end{figure}

The crystallization state of a Coulomb plasma is characterized by the ionic coupling parameter (see, e.g., \cite{ShapiroTeukolsky1983})
\begin{equation}\label{eq:Gamma}
\Gammac =
\frac{Z^{2} e^{2}}{a_{\mathrm{i}} k_{\mathrm{B}}T},
\qquad
a_{\mathrm{i}} =
\left(\frac{3 A m_{\mathrm{u}}}{4\pi\rho}\right)^{1/3},
\end{equation}
where $Z$ and $A$ are the ionic charge and mass number, respectively. For a classical one-component plasma, Monte Carlo and molecular-dynamics calculations place the freezing transition at $\Gammam\simeq 175$ \cite{FaroukiHamaguchi1993,PotekhinChabrier2000}.
Solving Eq.~(\ref{eq:Gamma}) for $\Gammac=\Gammam$ gives
\begin{equation}\label{eq:Tmelt}
\Tmelt =
\frac{Z^{2}e^{2}}{\Gammam k_{\mathrm{B}}}
\left(\frac{4\pi\rho}{3 A m_{\mathrm{u}}}\right)^{1/3}
= 7.1\times 10^6 \left(\frac{Z}{2}\right)^2\left(\frac{4}{A} \frac{\rho}{\rho_0}\right)^{1/3}\,\,\text{K},
\end{equation}
which is much lower than $T_0$. Indeed, from Eq. (\ref{eq:Gamma}), we obtain $\Gammac(T_0)\approx 0.62 \ll \Gammam$. The material is therefore better described as a non-crystallized weakly coupled ionic plasma. This conclusion is not sensitive to the assumed composition: for C$^{12}$ and O$^{16}$, 
$\Tmelt \sim \text{a few}\times 10^{8}$ K, and even for Fe$^{56}$, $\Tmelt\approx 5\times10^8\,\mathrm{K}\approx T_0/4$. Thus, the fiducial state is on the liquid side of the freezing boundary for all compositions considered here. Figure \ref{fig:phase_diagram} summarizes the result.

We now estimate the cooling timescale and compare it with that of the SSD. The thermal reservoir for the neutrino cooling is the heat capacity, $c_v$. The degenerate-electron thermal correction is smaller in the fiducial state, and the electron Fermi-sea energy is not a radiative heat reservoir for the local cooling. Thus, using the total specific internal energy instead of $c_v T$ would overestimate the cooling time. At fixed density, the dominant contribution is the ionic thermal heat capacity
\begin{equation}\label{eq:cv}
c_v^{\mathrm{ion}} =
\frac{3}{2}\frac{k_{\mathrm{B}}}{A m_{\mathrm{u}}}
\approx \left(\frac{4}{A}\right) 3.12\times10^{7}\,
\mathrm{erg\,g^{-1}\,K^{-1}}
\end{equation}
where $m_u$ is the atomic mass unit. At $(\rho_0,T_0)$, the dominant neutrino emission channel is the plasmon decay \cite{Haft1994}. The
corresponding emissivity is \cite{Itoh1996} $Q_{\mathrm{plasmon}} \approx
8.4\times10^{19}\,\mathrm{erg\,cm^{-3}\,s^{-1}}$, 
well above the photo-neutrino and pair-annihilation contributions under these conditions (see also, Appendix B in \cite{2016ApJ...833..107B}). The local cooling time may then be estimated as
\begin{equation}\label{eq:tcool_pre}
\tcool =
\frac{\rho\, c_v T}{Q_{\mathrm{tot}}(\rho,T)},
\end{equation}
which, for the fiducial state, gives $\tcool \approx 7.4\times10^6\,\mathrm{s}\approx 86$ d. 

The dynamical time associated with the turbulent amplification is many orders of magnitude shorter. In the simulations discussed below, the eddy turnover time is $\teddy \sim 5.0\times10^{-5}\,\mathrm{s}$, and the magnetic field reaches the nonlinear stage on millisecond timescales. Equivalently,
\begin{equation}\label{eq:timescomparison}
\begin{aligned}
\teddy &\ll \tdyn \ll \tcool,\\
\teddy &\sim 5\times10^{-5}\,\mathrm{s},\quad
\tdyn \sim 10^{-3}\,\mathrm{s},\quad
\tcool \sim 10^{7}\,\mathrm{s}.
\end{aligned}
\end{equation}
Thus, the cooling timescale exceeds the dynamo one by roughly ten orders of magnitude, and exceeds the eddy turnover time by more than eleven orders
of magnitude. We conclude that neutrino cooling does not modify the MHD dynamics during the simulated time evolution.

Thermal conduction does not alter the ordering (\ref{eq:timescomparison}) at the level required for the present argument. For a buried layer of thickness
$\Delta\sim10^{2}$--$10^{3}\,\mathrm{m}$ and crustal thermal diffusion coefficient
$\kappa_{\mathrm{th}}\sim10^{2}$--$10^{4}\,\mathrm{cm^2\,s^{-1}}$
\cite{Yakovlev2001,Potekhin2015}, one obtains
$t_{\mathrm{cond}}\sim\Delta^2/\kappa_{\mathrm{th}}
\sim10^{4}$--$10^{8}\,\mathrm{s}$, again far longer than the millisecond MHD evolution of the present problem. Residual compressional or nuclear heating, if present after fallback, would increase the thermal lifetime of the liquid layer rather than shorten it \cite{HaenselZdunik2008}.

The above calculation establishes the consistency of treating the local fiducial layer as non-crystallized during the MHD experiment. It does
not determine the global crystallization front, the detailed temperature profile of the newly accreted crust, or the composition produced by burning and electron captures. Those quantities depend on the fallback history, heat conduction, nuclear processing, and the coupling between the accreted envelope and the underlying NS. They should therefore be checked in a thermal-evolution calculation.

% ============================================================
\section{MHD model and numerical setup}
\label{sec:setup}
% ============================================================

\subsection{Equations and numerical method}
\label{sec:equations}

We solve the compressible resistive-MHD equations in a local Cartesian domain using \flash~4.7 \cite{Fryxell2000}. The simulations employ the
unsplit staggered-mesh (USM) solver \cite{LeeDeane2009,Lee2013}, with magnetic fields represented on staggered faces and advanced by
constrained transport. The equations below are written in the units used by the code, i.e., Gaussian-cgs units. The governing equations are
\begin{align}
\frac{\partial \rho}{\partial t}
   + \nabla\!\cdot(\rho\,\mathbf{v}) &= 0,
\label{eq:continuity}
\\[2pt]
\frac{\partial(\rho\mathbf{v})}{\partial t}
   + \nabla\!\cdot\!\bm{\Pi}
   &= \rho\,\mathbf{f}_{\mathrm{stir}},
\label{eq:momentum}
\\[2pt]
\frac{\partial \mathbf{B}}{\partial t}
   - \nabla\!\times\!(\mathbf{v}\!\times\!\mathbf{B})
   &= -\nabla\!\times\!\left(\eta\,\nabla\!\times\!\mathbf{B}\right),
\label{eq:induction}
\\[2pt]
\nabla\!\cdot\!\mathbf{B} &= 0,
\label{eq:divB}
\end{align}
with total energy evolved according to
\begin{multline}
\frac{\partial E}{\partial t}
   + \nabla\!\cdot\!\left[
      \left(E + P + \frac{1}{2}\mathbf{B}^{2}\right)\mathbf{v}
       - \mathbf{B}\left(\mathbf{v}\!\cdot\!\mathbf{B}\right)
      \right]
\\
   = \rho\,\mathbf{v}\!\cdot\!\mathbf{f}_{\mathrm{stir}}
      - Q_{\nu}(\rho,T)
      + \nabla\!\cdot\!\left[
        \eta\,\mathbf{B}\!\times\!
        \left(\nabla\!\times\!\mathbf{B}\right)\right].
\label{eq:energy}
\end{multline}
where
\begin{align}
\bm{\Pi} &=
\rho\,\mathbf{v}\mathbf{v}
       - \mathbf{B}\mathbf{B}
       + \left(P+\frac{1}{2}\mathbf{B}^{2}\right)\mathbb{I},
\label{eq:Pi}
\\
E &=
\rho e_{\mathrm{int}}
+ \frac{1}{2}\rho\,\mathbf{v}^{2}
+ \frac{1}{2}\mathbf{B}^{2}.
\label{eq:Etot}
\end{align}
The source terms are the imposed turbulent acceleration $\mathbf{f}_{\mathrm{stir}}$, the neutrino emissivity $Q_{\nu}$, and the explicit Ohmic resistivity $\eta$.

The pressure and specific internal energy are obtained from the Helmholtz EOS \cite{TimmesSwesty2000}, evaluated for our fiducial composition of a single fully ionized He$^4$ species with $A=4$, $Z=2$, and $Y_e=0.5$. In the EOS-control run P-C, the Helmholtz closure is replaced by an ideal gas with $P=(\gamma-1)\rho e_{\mathrm{int}}$ and $\gamma=5/3$.

The USM scheme is second-order accurate in space and time for smooth flows. We use the HLLD Riemann solver \cite{MiyoshiKusano2005}, the van Leer slope limiter with characteristic limiting, and an explicit Courant-Friedrichs-Lewy number ${\rm CFL}=0.4$, including the restriction from the resistive diffusion step. The divergence constraint is controlled by the constrained-transport update of the staggered magnetic field; in the periodic-box runs presented here, $\nabla\!\cdot\!\mathbf{B}$ remains at round-off level.

The simulations represent a local, non-stratified patch of the liquid post-hypercritical accretion layer. We adopt a fiducial thermodynamic state $(\rho_0,T_0)$, initial uniform magnetic field $\mathbf{B}_0 = B_0\,\hat{\mathbf{z}}$ with $B_0 = 10^{12}$ G, and initial velocity zero. The magnetic field strength is chosen to match a buried-strength seed field, but the initial geometry is not intended to represent a global buried configuration.

The computational domain is a cube of side
$L=100\,\mathrm{m}$ with periodic boundary conditions in all directions. The production runs use a uniform $128^3$ mesh, corresponding to $\Delta x=0.78\,\mathrm{m}$, while the resolution check P-A2 HR uses $256^3$ cells, corresponding to $\Delta x=0.39\,\mathrm{m}$.

We adopt a constant gravitational acceleration representative of the expected value at the NS surface: for $M = 1.4\,M_{\odot}$, $R= 10\,\mathrm{km}$ NS,
$g_{\mathrm{NS}} = G M/R^2 \simeq 1.86\times10^{14}\,\mathrm{cm\,s^{-2}}$. At the fiducial density and
electron fraction, the relativistic degenerate-electron pressure is approximately \cite{ShapiroTeukolsky1983}, $P_e \simeq
K_{\mathrm{rel}}(\rho Y_e)^{4/3}
\simeq 1.06\times10^{28}\,\mathrm{erg\,cm^{-3}}$, where $K_{\mathrm{rel}}=1.244\times10^{15}$ in cgs units, so $\Hp = P_e/(\rho\,g_{\mathrm{NS}})\approx  53$ m, and $L\simeq 1.9\Hp$. Therefore, the box is comparable to the local pressure scale height. The periodic setup should consequently be understood as a controlled turbulence experiment under post-hypercritical thermodynamic conditions. A self-consistent hydrostatic model accounting for a stratified crustal layer, buoyancy, vertical transport, compositional gradients, and a moving crystallization front is beyond the scope of the present simulations (see Sec.~\ref{sec:limitations}).

% ------------------------------------------------------------
\subsection{Source terms}
\label{sec:forcing}
% ------------------------------------------------------------

Turbulence is driven by an externally imposed stochastic acceleration field $\mathbf{f}_{\mathrm{stir}}(\mathbf{x},t)$. The forcing is constructed in Fourier space as an Ornstein-Uhlenbeck process \cite{EswaranPope1988,Schmidt2009,Federrath2010}. For each excited mode, the forcing amplitude evolves as
\begin{equation}
d\hat{\mathbf{f}}_{k}
=
-\frac{\hat{\mathbf{f}}_{k}}{\tau_{\mathrm{corr}}}\,dt
+
\sigma_k\,\mathcal{P}^{\perp}_{k}\!\cdot d\mathbf{W}_{t},
\label{eq:OU}
\end{equation}
where $\tau_{\mathrm{corr}}$ is the autocorrelation time, $d\mathbf{W}_t$ is a Wiener increment, and $\mathcal{P}^{\perp}_{ij}=\delta_{ij}-\hat{k}_i\hat{k}_j$ projects out the compressive component. Therefore, the forcing is, by construction, purely solenoidal, $\nabla\!\cdot\!\mathbf{f}_{\mathrm{stir}}=0$. 

The forcing is applied over the band $k_{\min}=6.28\times10^{-4}\,\mathrm{cm^{-1}}$, $k_{\max}=1.885\times10^{-3}\,\mathrm{cm^{-1}}$, corresponding to wavelengths between the box size and one-third of the box size. The parabolic forcing envelope peaks at $k_{\mathrm{peak}}=(k_{\min}+k_{\max})/2$, i.e. at an injection scale
$\lambda_{\mathrm{peak}}\simeq L/2$. The autocorrelation time is $\tau_{\mathrm{corr}}=10^{-4}\,\mathrm{s}$, comparable to the eddy turnover time in the reference run. The forcing normalization $\varepsilon_{\mathrm{st}}$ controls the resulting turbulent velocity and hence the magnetic Reynolds number. The values are listed in Table~\ref{tab:runs}.

As for the neutrino cooling term, we adopt a volumetric rate
\begin{equation}
Q_{\nu} =
Q_{\mathrm{pair}}
+ Q_{\mathrm{photo}}
+ Q_{\mathrm{plasmon}}
\label{eq:Qnu_total}
\end{equation}
that includes pair annihilation, photo-neutrino, and plasmon decay terms computed from the analytic fits in \cite{Itoh1996}. The high-density URCA cooling channel is inactive at the fiducial density. The cooling update solves $de_{\mathrm{int}}/dt=-Q_{\nu}/\rho$ and applies the corresponding change to the total energy. A temperature floor $T\ge10^7\,\mathrm{K}$ keeps the update within the validity range of the tabulated emissivity fits. In the control run P-B, cooling is switched off by setting $Q_{\nu}=0$.

We use a constant isotropic magnetic resistivity $
\eta = 10^{9}$ cm$^2$ s$^{-1}$. This value is not a physical estimate of the resistivity of the post-hypercritical layer. It is a numerical choice that keeps the resistive scale marginally resolved while placing the simulations in a dynamo-active regime. The magnetic Reynolds number is therefore much smaller than the physical value expected for a newborn NS envelope, but large enough to test SSD growth in a controlled resistive-MHD calculation. For the reference run, the resistive scale is resolved by only a few grid cells. This limitation is addressed through the $256^3$ resolution check and discussed further in Sec.~\ref{app:convergence}.

The Ohmic contribution is included consistently in the induction and energy equations. The resistive dissipation rate is
\begin{equation}
\dot{e}_{\mathrm{Ohm}}
=
\eta\,\left(\nabla\!\times\!\mathbf{B}\right)^2,
\label{eq:Q_ohm}
\end{equation}
and the dissipated magnetic energy is returned to the internal energy reservoir.

% ------------------------------------------------------------
\subsection{Control parameters}
\label{sec:diagnostics}
% ------------------------------------------------------------

The magnetic Reynolds number, Mach number, and effective magnetic Prandtl number are defined as
\begin{equation}\label{eq:dimless}
\Rm =
\frac{\vrms L}{\eta},
\qquad
\mathcal{M} =
\frac{\vrms}{c_s},
\qquad
\Pm =
\frac{\nu_{\mathrm{num}}}{\eta},
\end{equation}
where $\vrms=\langle\mathbf{v}^{2}\rangle^{1/2}$ is a volume average speed, $c_s$ is the sound speed, and $\nu_{\mathrm{num}}$ is the effective numerical viscosity of the USM solver. At the resolutions used here, $\nu_{\mathrm{num}}$ is comparable to the explicit magnetic diffusivity, so the simulations operate at an effective numerical $\Pm\sim 1$, set by the scheme and the adopted resistivity rather than by the microphysics. In the freshly accreted post-hypercritical layer modeled here, the electrons are strongly degenerate and the electrical conductivity is high, so the microscopic magnetic diffusivity $\eta=c^2/(4\pi\sigma)$ is many orders of magnitude smaller than the value adopted here, implying a microscopic magnetic Prandtl number $\Pm\gg 1$ \cite{Yakovlev2001,Potekhin2015}. Reaching this regime numerically is not feasible, so, as is standard in dynamo and MRI simulations of compact-object plasmas, an order-unity value is adopted as a compromise \cite{ThompsonDuncan1993,ReboulSalze2021}. Thus, our conclusions are restricted to the resolved, numerical-$\Pm$ regime explored in the simulations.

We monitor the volume-averaged kinetic and magnetic energies, $\Ekin$ and $\Emag$, the rms magnetic field $\Brms$, the turbulent Mach
number, and the magnetic-to-kinetic energy ratio
$\fsat=\Emag/\Ekin$ in the saturated phase. Growth rates are measured from the exponential phase of $\Emag(t)$ before nonlinear magnetic back-reaction becomes important. We use $\Emag/\Ekin=0.05$ as indicative of the onset of nonlinear
back-reaction. This threshold is used only to compare runs consistently, not as a universal saturation criterion.

Table~\ref{tab:runs} summarizes the values of the parameters of the six simulations used in this work. Runs P-A1, P-A2, and P-A3 form the $128^3$ magnetic-Reynolds-number survey with identical microphysics and different forcing amplitudes. Runs P-A2, P-B, and P-C form the physics-control set: P-A2 and P-B isolate the effect of neutrino cooling, while P-B and P-C isolate the effect of the thermodynamic closure. Run P-A2 HR is the $256^3$ counterpart of P-A2 and provides a resolution check for the reference case.

\begin{table}[!ht]
\centering
\caption{Production run matrix and key measurements. $\varepsilon_{\mathrm{st}}$ is the stochastic-forcing normalization; $\Rm=\vrms L/\eta$ is the box-scale magnetic Reynolds number;
$\mathcal{M}=\vrms/c_s$ is the turbulent Mach number; $\Bsat/B_0$ is the saturation amplification factor; and $\fsat=\Emag/\Ekin$ is the saturated magnetic-to-kinetic energy ratio.}
\label{tab:runs}
\setlength{\tabcolsep}{4pt}
\begin{tabular}{@{}llcccccc@{}}
\hline\hline
Run    & EOS    & Cool. & $\varepsilon_{\mathrm{st}}$ & $\Rm$ & $\mathcal{M}$ & $\Bsat/B_0$ & $\fsat$ \\
\hline
P-A1    & Helm. & ON  & $10^{17}$ &  670  & 0.06 &   8.4 & 0.13 \\
P-A2    & Helm. & ON  & $10^{19}$ & 2150  & 0.21 &  34.5 & 0.20 \\
P-A2 HR & Helm. & ON  & $10^{19}$ & 2280  & 0.22 &  35.5 & 0.21 \\
P-A3    & Helm. & ON  & $10^{20}$ & 3700  & 0.37 &  68.4 & 0.28 \\
\hline
P-B     & Helm. & OFF & $10^{19}$ & 2050  & 0.20 &  39.5 & 0.30 \\
P-C     & Ideal & OFF & $10^{19}$ & 2100  & 0.19 &  40.4 & 0.35 \\
\hline
\end{tabular}
\end{table}

% ============================================================
\section{Results}
\label{sec:results}
% ===========================================================
\begin{figure*}[t!]
\centering
\includegraphics[width=\textwidth]{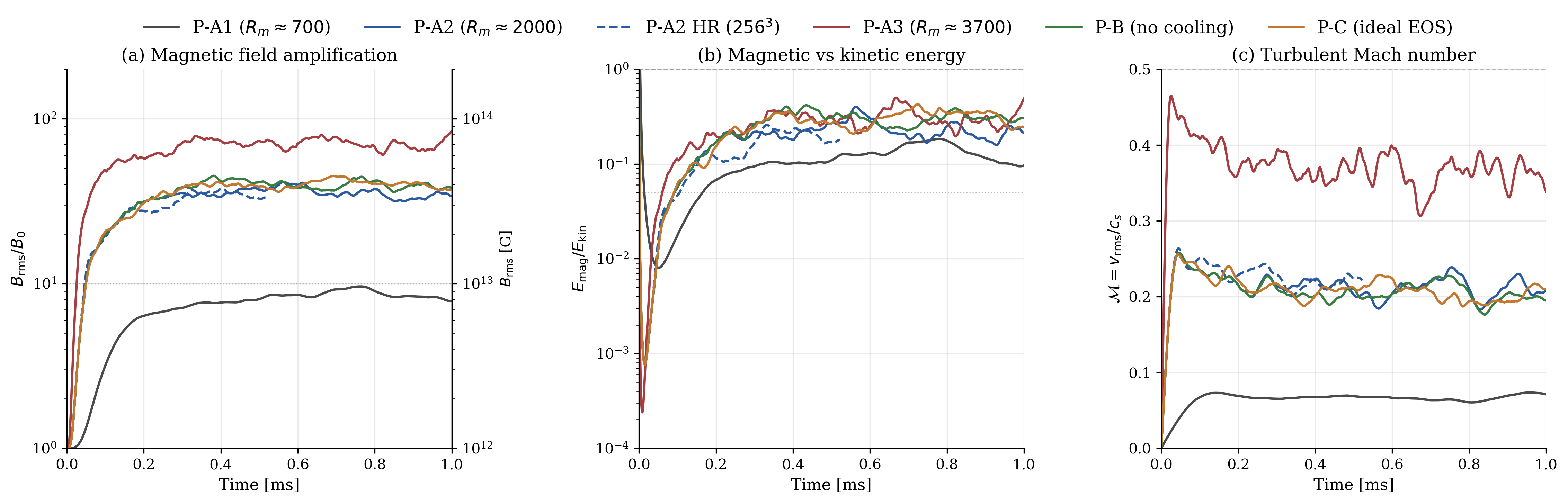}
\caption{Global evolution of the six simulations.
\textbf{(a)} Magnetic-field amplification, $\Brms/B_0$. The right axis gives $\Brms$ in physical units. \textbf{(b)} Magnetic-to-kinetic energy ratio, $\Emag/\Ekin$. \textbf{(c)} Turbulent Mach number, $\mathcal{M}=\vrms/c_s$. The runs with $\Rm\gtrsim2000$ (P-A2, P-A2 HR, P-B, and P-C) saturate at $\Brms/B_0\simeq35$--$40$, while the highest-$\Rm$ case P-A3 reaches $\Brms/B_0\simeq70$. The lowest-$\Rm$ case P-A1 shows weaker amplification and saturates at $\Brms/B_0\simeq8$. The dotted horizontal line in panel (b) marks the threshold $\Emag/\Ekin=0.05$, used here to identify the onset of nonlinear magnetic back-reaction.}
\label{fig:unified}
\end{figure*}

Figure~\ref{fig:unified} shows that in all six runs, the rms magnetic field grows from the seed value
$B_0=10^{12}$ G and reaches a statistically steady saturated level, $\Bsat$.
Panel~(a) shows the amplification factor $\Brms/B_0$. There is a monotonic increase in the saturation amplitude with the Reynolds number, hence with increasing forcing strength. Run P-A1, with $\Rm\simeq670$, reaches $\Brms/B_0\simeq8$. Run P-A2, with $\Rm\simeq2150$, reaches $\Brms/B_0\simeq35$, corresponding to $\Bsat\simeq3.5\times10^{13}\,\mathrm{G}$. Run P-A3, with $\Rm\simeq3700$, reaches $\Brms/B_0\simeq70$, or $\Bsat\simeq7\times10^{13}\,\mathrm{G}$. Because the run suite does not include a subcritical case, these simulations do not determine $\Rmcrit$, but show that for the range $\Rm\simeq700$--$3700$, the system lies in a dynamo-active regime with the saturation level increasing with $\Rm$. The runs P-B and P-C, both close to the reference forcing amplitude, reach $\Brms/B_0\simeq 40$. Their saturation values are slightly higher than that of P-A2, but the difference is modest compared with the change between the low-, reference-, and high-$\Rm$ cases. This behavior is consistent with small differences in the measured turbulent velocities and with realization-to-realization scatter in the saturated state (see Sec.~\ref{app:convergence}).

Panel~(b) shows the ratio $\Emag/\Ekin$. At the beginning of the run, the velocity field is still being established by the stochastic forcing, so
the initial value of this ratio is not physically meaningful. After a short transient, the kinetic energy reaches the forced-turbulence level, and the magnetic energy enters an exponential growth phase. Nonlinear magnetic back-reaction becomes visible when $\Emag/\Ekin$ approaches a few percent. 
The saturated states are in sub-equipartition. We measure $\fsat=\Emag/\Ekin\simeq0.13$ for P-A1,
$\fsat\simeq0.20$--$0.21$ for P-A2 and P-A2 HR,
$\fsat\simeq0.28$ for P-A3, $\fsat\simeq0.30$ for P-B, and $\fsat\simeq0.35$ for P-C. These values are consistent with the range commonly found in numerical studies of subsonic SSDs at numerical $\Pm\sim1$ \cite{HaugenBrandenburgDobler2004,Schekochihin2004ApJ,Federrath2011,SetaFederrath2020,SetaFederrath2022}.

Panel~(c) shows that all simulations remain subsonic. The four runs with the reference forcing amplitude, P-A2, P-A2 HR, P-B, and P-C, cluster at
$\mathcal{M}\simeq0.19$--$0.22$. The low- and high-forcing runs bracket this range, with $\mathcal{M}\simeq 0.06$ for P-A1 and $\mathcal{M}\simeq 0.37$ for P-A3. The small Mach numbers explain why density fluctuations remain weak and why the induction equation is controlled primarily by the velocity field and the magnetic diffusivity, rather than by compressible shocks or large thermodynamic fluctuations.

% ============================================================
% Wide figure: morphological evolution
% ============================================================
\begin{figure*}[t!]
\centering
\includegraphics[width=\textwidth]{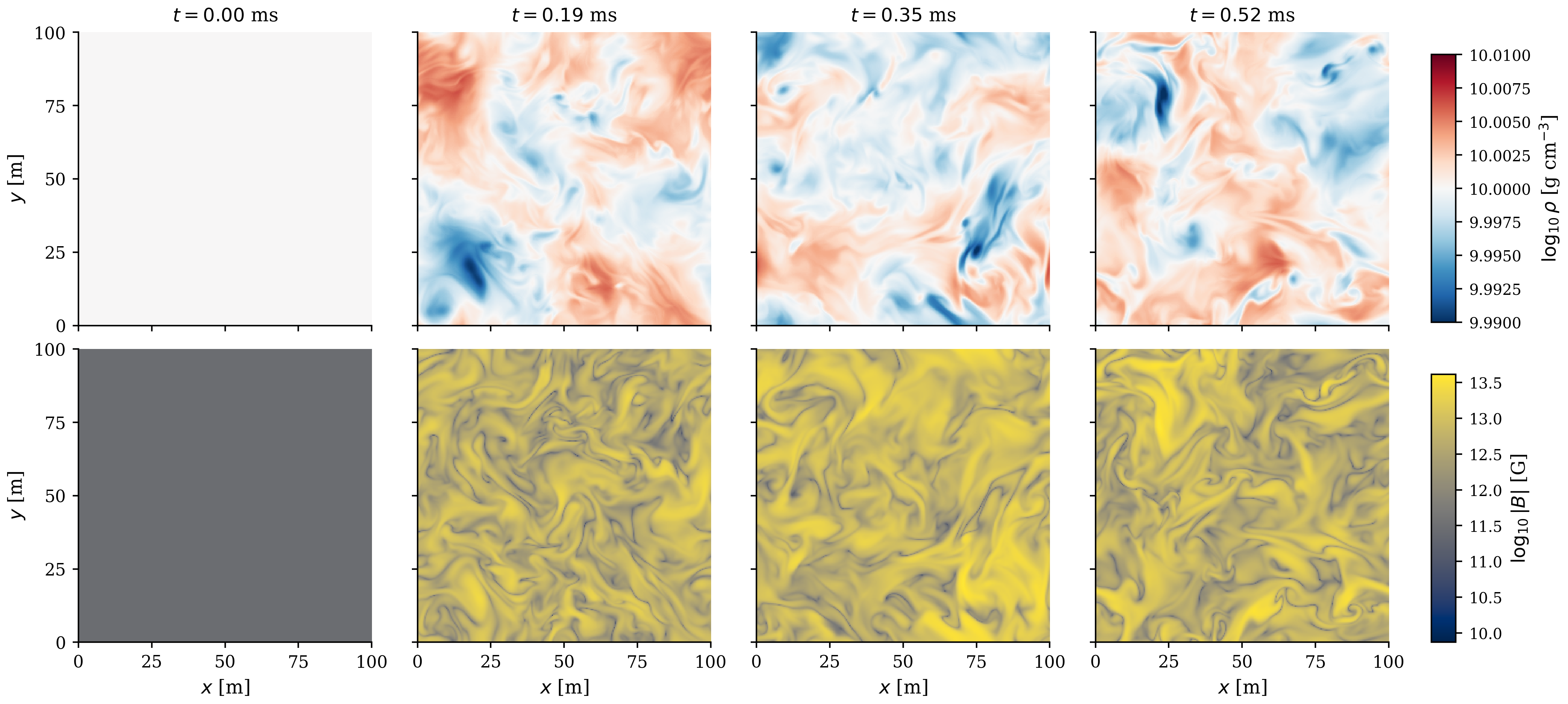}
\caption{Morphological evolution of the reference run P-A2, shown as two-dimensional slices through the box midplane ($z=L/2$) at four selected times. \emph{Top row:} $\log_{10}\rho$  [g\,cm$^{-3}$], displayed with a diverging colormap centered on the mean density $\rho_0=10^{10}\,\mathrm{g\,cm^{-3}}$ to 
highlight the small compressive fluctuations. \emph{Bottom row:} $\log_{10}|B|$ [G], with a sequential colormap spanning the seed-field value $B_0=10^{12}\,\mathrm{G}$ and the amplified 
state. From left to right: initial condition ($t=0$), kinematic  phase ($t=0.19\,\mathrm{ms}$), nonlinear onset ($t=0.35\,\mathrm{ms}$), and saturated state  ($t=0.52\,\mathrm{ms}$). The initially uniform seed field is stretched and folded into intermittent sheets and filaments characteristic of the SSD action. Density fluctuations remain small, $\delta\rho/\rho\sim10^{-4}$, consistent with the subsonic, nearly incompressible regime.}
\label{fig:snapshots}
\end{figure*}

% ============================================================
% Wide figure: spectral signature
% ============================================================
\begin{figure*}[t!]
\centering
\includegraphics[width=\textwidth]{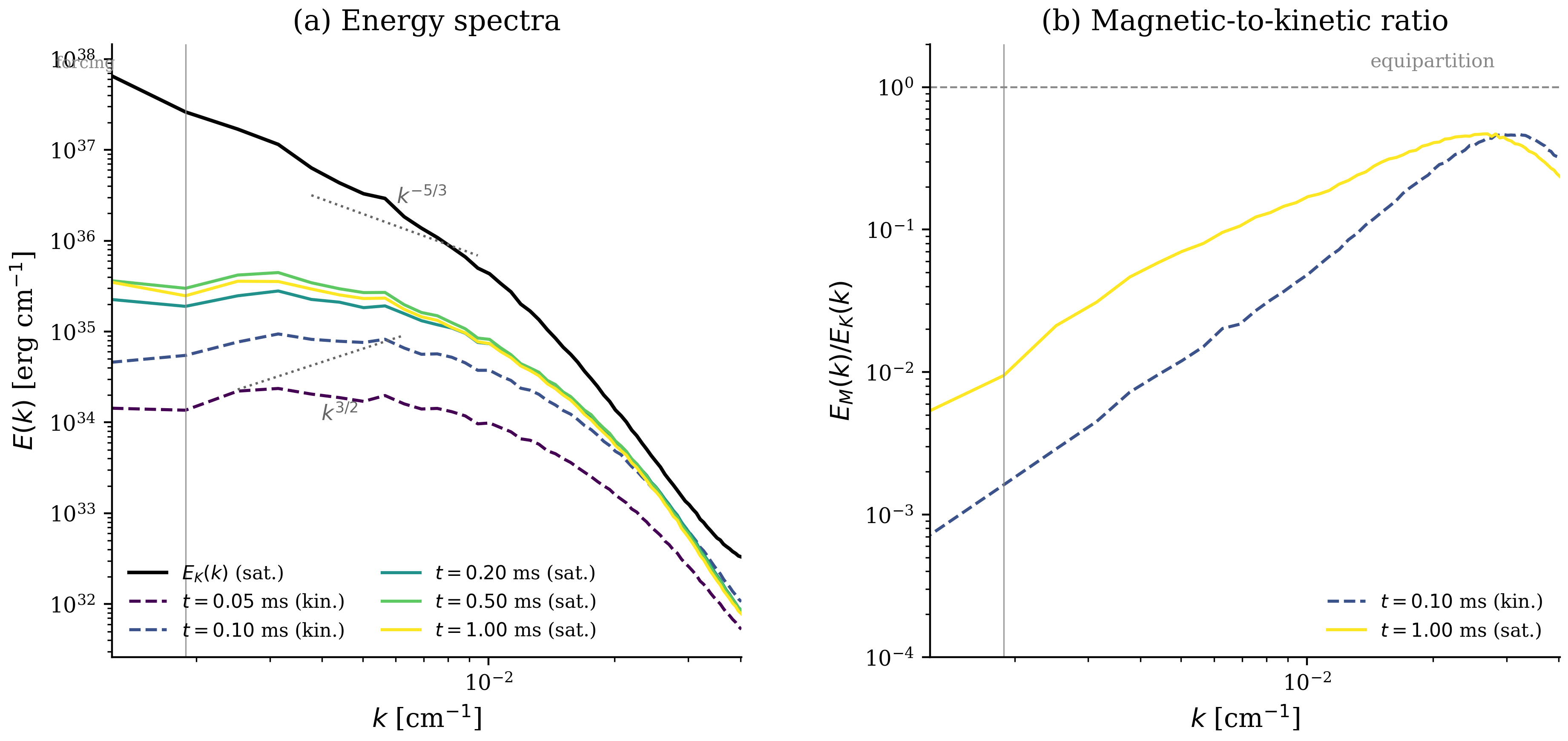}
\caption{Spectral signature of the SSD in run P-A2. \textbf{(a)} Kinetic energy spectrum $E_K(k)$ at saturation (solid black) and magnetic energy spectra $E_M(k)$ at five selected times spanning the kinematic phase ($t = 0.05, 0.10\,\mathrm{ms}$, dashed) and the saturated phase ($t = 0.20, 0.50, 1.00\,\mathrm{ms}$, solid). Dotted lines indicate the Kazantsev $k^{3/2}$ slope characteristic of kinematic SSD growth and the Kolmogorov $k^{-5/3}$ inertial-range slope. The vertical grey line marks the forcing scale $k_{\mathrm{f}} \simeq 2\pi/L$. \textbf{(b)} Ratio $E_M(k)/E_K(k)$ during the kinematic phase ($t = 0.10\,\mathrm{ms}$) and at saturation ($t = 1.00\,\mathrm{ms}$). The horizontal dashed line indicates spectral equipartition.}
\label{fig:spectra}
\end{figure*}

We turn to the morphology of the magnetic field and the density. Figure~\ref{fig:snapshots} shows midplane slices of the density and magnetic-field strength for the reference run P-A2. The magnetic field evolves from a uniform seed into a tangled, intermittent structure. During the kinematic stage, the flow stretches the field into narrow sheets and filaments. By $t\simeq0.4$--$0.5\,\mathrm{ms}$ the magnetic structure is fully developed, and the run has entered the nonlinear saturated regime. The density field, by contrast, remains nearly uniform: fluctuations are of order $\delta\rho/\rho\sim10^{-4}$, consistent with the low Mach number. This weak compressibility indicates that magnetic growth is driven by turbulent stretching and folding of field lines rather than by compression, consistent with a local SSD at work.

The spectral signature of the dynamo provides the most stringent test of its small-scale character. Figure~\ref{fig:spectra} shows the kinetic [Panel (a)] and magnetic [Panel (b)] energy spectra of run P-A2 at five representative times. During the kinematic phase ($t \lesssim 0.15\,\mathrm{ms}$), the magnetic spectrum $E_M(k)$ peaks at scales well below the forcing scale and exhibits a positive slope in the large-scale range consistent with the Kazantsev $k^{3/2}$ prediction \cite{Kazantsev1968,Schekochihin2004ApJ}, characteristic of a stochastically driven SSD. The kinetic spectrum $E_K(k)$ displays a well-developed inertial range with a slope compatible with the Kolmogorov $k^{-5/3}$ scaling, confirming that the turbulent cascade is properly resolved at the simulation resolution.

As the dynamo enters saturation, the magnetic spectrum amplifies by more than three orders of magnitude across all scales, but its shape changes: the peak shifts slightly toward smaller scales, and the spectrum flattens at large $k$, indicating saturation due to back-reaction on the smallest turbulent eddies. Crucially, the saturated $E_M(k)$ remains sub-equipartition at large and intermediate scales, with $E_M(k)/E_K(k)\sim 10^{-2}$--$10^{-1}$, and approaches equipartition only near the resistive scale. This residual deficit of large-scale magnetic energy is the canonical SSD signature reported in incompressible 
\cite{Schekochihin2004ApJ,BrandenburgSubramanian2005}, and compressible \cite{Federrath2011,SetaFederrath2020,Achikanath2021} simulations, and confirms that the field amplification in our setup is driven by the SSD mechanism.

We measure the kinematic growth rate from the exponential rise of the magnetic energy. Since $\Emag\propto B^2$, we write $\Emag(t) \propto \exp(2\gamma t)$, where $\gamma$ is the magnetic-field growth rate. The fitting interval starts after the initial forcing transient, at the global minimum of
$\Emag(t)$, and ends when $\Emag/\Ekin=0.05$. This choice avoids both the artificial start-up phase and the nonlinear saturation stage. For the reference run P-A2, we obtain $\gamma\teddy\simeq2.6$, where $\teddy=L/\vrms$ is the box-scale eddy turnover time. The high-resolution run P-A2 HR gives a value within $\simeq7\%$ of the reference result (Table~\ref{tab:convergence}). 
The measured value falls within the broad range reported in numerical SSD studies at comparable numerical $\Pm$ and moderate $\Rm$
\cite{Federrath2011,Federrath2014,SetaFederrath2020,Achikanath2021}.
The saturation field can be understood from a simple energy argument. In the saturated state, the magnetic energy reaches a fraction $\fsat$ of the turbulent kinetic energy. Expressed in physical Gaussian units, this gives
\begin{equation}\label{eq:Bsat}
\Bsat = \vrms\sqrt{4\pi\,\fsat\,\rho_0}.
\end{equation}
For P-A2, using $\vrms\simeq2.1\times10^8\,\mathrm{cm\,s^{-1}}$,
$\rho_0=10^{10}\,\mathrm{g\,cm^{-3}}$, and
$\fsat\simeq0.20$, Eq.~(\ref{eq:Bsat}) gives
$\Bsat\simeq3.3\times10^{13}\,\mathrm{G}$, in agreement with the measured value $\simeq3.5\times10^{13}\,\mathrm{G}$. For P-A3, using $\vrms\simeq3.7\times10^8\,\mathrm{cm\,s^{-1}}$ and $\fsat\simeq0.28$, the same estimate gives
$\Bsat\simeq6.9\times10^{13}\,\mathrm{G}$, again consistent with the numerical simulation.

This scaling indicates that once the SSD reaches saturation, $\Bsat$ is set primarily by the turbulent kinetic energy reservoir and the density, and is independent of the initial seed amplitude. The role of $B_0$ is to set the duration of the kinematic phase. For a weaker seed, the same saturated state would
be reached later, provided the forcing persists long enough.

% ------------------------------------------------------------
\section{Limitations}
\label{sec:limitations}
% ------------------------------------------------------------

Before concluding, we summarize and discuss the main limitations that circumscribe the validity of the results of the numerical simulations presented in this article.

\paragraph{Sustained versus decaying turbulence}
The simulations use continuous Ornstein-Uhlenbeck forcing to obtain growth rates and saturation levels under statistically controlled conditions. In the post-hypercritical accretion layer, turbulence inherited from the accretion envelope may decay after fallback subsides. In this context, our saturated amplitudes should be interpreted as the outcome if turbulent motions persist for several eddy turnover times. A decaying-turbulence calculation is needed to determine whether the kinetic-energy reservoir is sufficient to reach the same saturation level.

\paragraph{Forcing geometry}
The forcing is purely solenoidal, which favors SSD growth relative to compressive forcing \cite{Federrath2011,Federrath2014}. Real post-fallback turbulence is unlikely to be purely solenoidal. Mixed or more compressive forcing could reduce the growth rate and possibly the saturation fraction at fixed $\Rm$. 

\paragraph{Resolution and resistive scale}
The $128^3$ runs marginally resolve the resistive scale. The $256^3$ reference run shows that $\Brms/B_0$, $\fsat$, $\vrms$, and $\mathcal{M}$ are robust to a factor-of-two increase in linear resolution, but this is not a formal convergence study of the turbulent spectrum or of the dissipation-range structure. Higher-resolution calculations are required to measure the growth rate and magnetic spectra with greater precision.

\paragraph{Magnetic Prandtl number}
The simulations operate at an effective numerical $\Pm\sim1$ because the explicit magnetic diffusivity is comparable to the scheme's effective numerical viscosity. The microscopic transport coefficients in a hot newborn NS envelope are uncertain and vary with density, temperature, and composition. Our conclusions are restricted to the resolved finite-$\Rm$, numerical-$\Pm$ regime studied here. 

\paragraph{Periodic geometry and stratification} The domain is periodic, uniform, and has constant gravitational acceleration. While turbulent stretching is captured, hydrostatic stratification, buoyancy, vertical transport, compositional gradients, and the motion of a crystallization front are not considered. This idealization could be significant as the box size is comparable to the local pressure-gradient scale height.

\paragraph{Initial magnetic geometry}
The seed magnetic field is uniform. This is adequate for a controlled SSD test, but is probably not representative of a realistic buried configuration. Multidimensional MHD simulations of magnetic submergence by hypercritical fallback find that the advected field is left with a complicated, predominantly non-dipolar geometry, sheared and concentrated near the interface between the newly accreted material and the underlying pre-existing layers \cite{GeppertPageZannias1999,Bernal2013,TorresForne2016}. Such a configuration could modify both the onset of local amplification and the subsequent magnetic topology.

\paragraph{Composition and thermal evolution}
The fiducial model uses He$^4$ ($Z=2$, $A=4$). The post-hypercritical accretion layer may contain heavier ashes and evolve through burning, electron capture, and compositional separation. These effects can change the EOS, conductivities, neutrino emissivities, and freezing temperature. The composition considered in Sec.~\ref{sec:crystallization} shows that the fiducial state remains liquid for representative species, but does not replace a full compositional-evolution calculation.

\paragraph{Connection to observable magnetar fields}
The simulations amplify the local rms magnetic field to $\sim10^{13}$--$10^{14}$ G. They do not show that this energy becomes an observable external dipole, nor that it produces a magnetar. The calculation constrains a possible local source of internal small-scale magnetic energy, but its subsequent conversion, survival, or reorganization belongs to the solid-crust Hall-Ohmic stage.

% ============================================================
\section{Conclusions}
\label{sec:conclusions}
% ============================================================

We have carried out a local 3D resistive-MHD study of SSD action under thermodynamic conditions motivated by the liquid post-hypercritical accretion layer of a newborn NS. The calculations are designed as a controlled viability test of SSD action under post-hypercritical accretion conditions. Within this scope and the limitations discussed in Sec. \ref{sec:limitations}, our main conclusions are as follows.

\begin{enumerate}
\item The fiducial thermodynamic state,
$\rho_0=10^{10}\,\mathrm{g\,cm^{-3}}$ and
$T_0=2\times10^9\,\mathrm{K}$, lies well on the liquid side of the Coulomb freezing boundary. The local cooling time is many orders of magnitude longer than the eddy turnover and magnetic-amplification times. This supports the use of fluid MHD to test local magnetic amplification during early post-hypercritical accretion.

\item Externally forced subsonic turbulence drives exponential magnetic growth in all simulations with $\Rm\simeq700$--$3700$. Starting from
$B_0=10^{12}\,\mathrm{G}$, the rms magnetic field saturates at $\Bsat\simeq3\times10^{13}$--$7\times10^{13}\,\mathrm{G}$ in the reference and high-$\Rm$ cases, corresponding to amplification factors of $\sim35$--$70$ on millisecond timescales.

\item The saturated states remain sub-equipartition, with $\fsat=\Emag/\Ekin\simeq0.2$--$0.3$ for the main dynamo-active runs. The saturation fields obey the scaling (\ref{eq:Bsat}), indicating that the final field strength is determined primarily by the turbulent kinetic energy reservoir and the density, independent of the seed-field amplitude.

\item The control runs show that, in the subsonic regime explored here, neutrino cooling has no measurable dynamical effect on the magnetic growth over the simulated interval, and the choice between the Helmholtz and ideal-gas closures weakly affects the results. The $256^3$ reference run confirms that $\Brms/B_0$, $\fsat$, $\vrms$, and $\mathcal{M}$ are robust to a factor-of-two increase in linear resolution.
\end{enumerate}

The field strengths reached in the saturated runs,
$\Bsat\sim (3$--$7)\times10^{13}$ G, are comparable to the internal multipolar fields commonly invoked in hidden-field models of CCOs and low-field magnetars \cite{ShabaltasLai2012,Bogdanov2014,Luo2015,Igoshev2021CCO}. These results indicate that SSD action in the liquid post-hypercritical layer may provide an additional source of small-scale magnetic energy prior to the onset of Hall-Ohmic evolution. If even a modest fraction of this energy survives, it could contribute to the hidden magnetic reservoir inferred in CCOs and low-field magnetars. Further simulations of the subsequent Hall-Ohmic evolution are needed to determine how much of this magnetic energy survives crystallization, how it couples to Hall drift in the solid crust, or how it reorganizes into a large-scale component capable of affecting the external dipole.

%============================================================
% Acknowledgements
% %%%%%%%%%%%%%%%%%%%%%%%%%%%%%%%%%%%%%%%%%%%%%%%%%%%%%
\section*{Acknowledgments}

D.F.B. acknowledges the support of the Observatorio Astron\'omico Nacional, Universidad Nacional de Colombia, and the fellowship support
received during his M.Sc. studies in Astronomy. C.G.B. acknowledges support from the Universidad Nacional de Colombia,
where he is a faculty member in physics and astrophysics at the Departamento de F\'isica, Facultad de Ciencias.
J.A.R. acknowledges support from the Università degli Studi di Ferrara and ICRANet at the University of Ferrara.
The authors acknowledge the FLASH Center for Computational Science at the University of Rochester for making the \flash code available to the astrophysical community. The simulations presented in this work were performed on the high-performance computing facilities of the Centre for Analytics, Informatics and Research (CAIR) at Memorial University of Newfoundland, Canada, whose computational resources and technical support are gratefully acknowledged.

\bibliographystyle{elsarticle-num} %elsarticle-harv
\bibliography{references}

\appendix

% ------------------------------------------------------------
\section{Resolution and microphysical checks}
\label{app:convergence}
% ------------------------------------------------------------

We simulate the reference case P-A2 at $256^3$ cells to assess the sensitivity of the main simulation parameters to resolution. The comparison with the $128^3$ run is shown in Table~\ref{tab:convergence}.

\begin{table}[!ht]
\centering
\caption{Resolution check for the reference model: P-A2 ($128^3$) versus P-A2 HR ($256^3$). The quoted difference is relative to the $128^3$ value.}
\label{tab:convergence}
\begin{tabular}{@{}lccc@{}}
\hline\hline
Metric             & $128^3$  & $256^3$  & $\Delta$ \\
\hline
$\Rm$              &  2150    &  2280    & $+5.9\,\%$ \\
$\mathcal{M}$      &  0.213   &  0.218   & $+2.3\,\%$ \\
$\vrms$ [cm\,s$^{-1}$] & $2.13\times 10^8$ & $2.18\times 10^8$ & $+2.3\,\%$ \\
$\gamma\,\teddy$   &  2.58    &  2.76    & $+7.0\,\%$ \\
$\Bsat$ [G]        & $3.45\times 10^{13}$ & $3.55\times 10^{13}$ & $+2.9\,\%$ \\
$\Bsat / B_0$      &  34.5    &  35.5    & $+2.9\,\%$ \\
$\fsat$            &  0.203   &  0.213   & $+4.5\,\%$ \\
\hline
\end{tabular}
\end{table}

The two resolutions agree at the few-percent level in $\vrms$, $\mathcal{M}$, $\Bsat/B_0$, and $\fsat$, while the fitted growth rate differs by about $7\%$. Thus, we can consider that the obtained values are robust to this factor-of-two increase in linear resolution. A detailed spectral convergence is left for a future higher-resolution work.

Regarding the microphysics, the runs assess two choices: neutrino cooling and the thermodynamic EOS. P-A2 and P-B differ by the presence or absence of the neutrino-cooling source term. P-B and P-C both omit cooling but use the Helmholtz and ideal-gas EOS, respectively. Runs P-A2 and P-B show very similar magnetic growth and saturation. The measured saturation factors are $\Bsat/B_0 = 34.5$ and $39.5$, respectively. This difference is modest compared with the variation across the $\Rm$ survey and is consistent with the slightly different measured turbulent velocities and the scatter in saturation levels. More importantly, no qualitative change appears in the growth curve when cooling is removed.

This behavior is expected from the timescale estimate [see Eq. (\ref{eq:timescomparison})]. At the fiducial thermodynamic state $(\rho_0,T_0)$, the neutrino cooling time is many orders of magnitude longer than the eddy turnover and magnetic-amplification times. During the simulated millisecond interval, turbulent driving and dissipation dominate the energy evolution. Thus, in the regime explored here, cooling sets the long-term thermal lifetime of the liquid layer but does not affect the short-time SSD growth.

Runs P-B and P-C produce almost identical saturated amplification factors, $\Bsat/B_0=39.5$ and $40.4$, and similar Mach numbers, $\mathcal{M}\approx 0.20$ and $0.19$. This result does not imply that the EOS is generally unimportant in newborn NS envelopes. It indicates that, for the present conditions (subsonic, nearly incompressible, solenoidal forced flow), the SSD is controlled mainly by $\Rm$ and the velocity field. The weak EOS sensitivity agrees with the small density fluctuations seen in Fig.~\ref{fig:snapshots}. For weak compressibility, the induction equation is governed by stretching, folding, and diffusion of the magnetic field by the velocity field. Thermodynamic differences could be more important in regimes with stronger compressibility, shocks, buoyancy, stratification, or larger temperature and composition contrasts, which are absent from the present setup.

\end{document}